\documentclass{article}
\usepackage[T1]{fontenc} 
\usepackage[utf8]{inputenc} 
\usepackage{ismir,amsmath,cite,url}
\usepackage{graphicx}
\usepackage{color}
\usepackage{amsfonts}
\usepackage{dsfont} 
\usepackage{fixltx2e}
\usepackage{multirow}
\usepackage{comment}
\usepackage{makecell} 
\usepackage{nicefrac}

\usepackage{booktabs}
\usepackage[hidelinks]{hyperref}
\usepackage{lineno}

\makeatletter
\newcommand*{\getcountref}[1]{%
\expandafter\@getcountref\csname r@#1\endcsname
}
\newcommand*{\@getcountref}[1]{%
\ifx#1\relax
0
\else
\expandafter\@car#1\@empty\@nil
\fi
}
\makeatother

\title{Supervised and Unsupervised Learning of Audio Representations for Music Understanding}

\multauthor
{Matthew C. McCallum \hspace{1cm} Filip Korzeniowski \hspace{1cm} Sergio Oramas} { \bfseries{Fabien Gouyon \hspace{1cm} Andreas F. Ehmann}\\
  SiriusXM, USA
}

\def\authorname{Matthew C. McCallum, Filip Korzeniowski, Sergio Oramas, Fabien Gouyon, Andreas F. Ehmann}

\begin{document}

\maketitle
\begin{abstract}
In this work, we provide a broad comparative analysis of strategies for pre-training audio understanding models for several tasks in the music domain, including labelling of genre, era, origin, mood, instrumentation, key, pitch, vocal characteristics, tempo and sonority. Specifically, we explore how the domain of pre-training datasets (music or generic audio) and the pre-training methodology (supervised or unsupervised) affects the adequacy of the resulting audio embeddings for downstream tasks.

We show that models trained via supervised learning on large-scale expert-annotated music datasets achieve state-of-the-art performance in a wide range of music labelling tasks, each with novel content and vocabularies. This can be done in an efficient manner with models containing less than 100 million parameters that require no fine-tuning or reparameterization for downstream tasks, making this approach practical for industry-scale audio catalogs.

Within the class of unsupervised learning strategies, we show that the domain of the training dataset can significantly impact the performance of representations learned by the model. We find that restricting the domain of the pre-training dataset to music allows for training with smaller batch sizes while achieving state-of-the-art in unsupervised learning---and in some cases, supervised learning---for music understanding.

We also corroborate that, while achieving state-of-the-art performance on many tasks, supervised learning can cause models to specialize to the supervised information provided, somewhat compromising a model's generality.

\end{abstract}
\section{Introduction}\label{sec:introduction}

In this work, we consider a broad array of classification and labelling tasks under the umbrella of music understanding. Such tasks include the labelling of genre, origin, mood, musical key, instruments, era, emotion and pitch present in music. These tasks have many applications in industry, particularly in music streaming and recommendation services where automated understanding of audio can assist in a range of tasks such as organizing, filtering and personalizing content to a listener's taste and context.

Recent research in automated audio understanding has focused on training convolutional \cite{kazakos2021slow, wang2022towards, pons2018end, wang2021multimodal, jansen2018unsupervised, kong2020panns, tagliasacchi2020pre, saeed2021contrastive, wang2021multi, spijkervet2021contrastive, lee2018samplecnn, huang2020large} and / or attention based \cite{gong2021ast, lu2021spectnt, won2021semi, yan2019a, liu2020mockingjay, gong2021psla, baevski2020wav2vec, koutini2021efficient, manco2022learning} networks on moderately large collections of frequency-domain audio \cite{jansen2018unsupervised, gong2021ast, kong2020panns, tagliasacchi2020pre, saeed2021contrastive, wang2022towards, lu2021spectnt, huang2020large, gong2021psla, koutini2021efficient, manco2022learning}, time-domain audio \cite{pons2018end, spijkervet2021contrastive, lee2018samplecnn, baevski2020wav2vec} or multi-format / multi-modal \cite{wang2021multimodal, wang2021multi, won2021multimodal, manco2022learning} data. Such models are often trained on tags encompassing some of the musical labels listed above, achieving promising results \cite{lee2018samplecnn, kazakos2021slow, pons2018end, gong2021ast, kong2020panns, lu2021spectnt, won2021semi, won2021multimodal}. More recent works propose unsupervised strategies for music understanding such as contrastive learning \cite{wang2022towards, wang2021multimodal, jansen2018unsupervised, saeed2021contrastive, wang2021multi, spijkervet2021contrastive, manco2022learning} or predictive / generative approaches \cite{tagliasacchi2020pre, dhariwal2020jukebox, manco2022learning, koutini2021efficient}. Unsupervised strategies are appealing because they require no annotated data and generalize well to new tasks \cite{wang2022towards, manco2022learning}, but lag the performance of supervised learning at a similar scale \cite{wang2022towards, spijkervet2021contrastive}. Generative learning strategies \cite{tagliasacchi2020pre, dhariwal2020jukebox} have been shown to achieve competitive, and sometimes state-of-the-art (SOTA), performance in several music understanding tasks \cite{castellon2021codified}, although, currently there is no evaluation demonstrating the effectiveness of this approach to any of the aforementioned approaches, at comparable scale.

Modern music streaming services have very large music catalogs that amount to many petabytes of audio data if uncompressed. Due to the scale of this data it is desirable to build models that are efficiently scalable, and understand audio in a general enough way that, as needs or requirements change, they may be used to solve novel problems without reprocessing such data. Models in the order of 10M or 100M parameters are currently relatively cost-effective to both train and apply inference to industry-scale catalogs, whilst models consisting of billions of parameters, e.g.\, that evaluated in~\cite{castellon2021codified}, are typically impractical, or very expensive, for both training and inference.

More recently, research has adopted approaches producing generalized audio embeddings~\cite{wang2022towards, wang2021multimodal, jansen2018unsupervised, kong2020panns, tagliasacchi2020pre, saeed2021contrastive, wang2021multi, spijkervet2021contrastive, kim2020one, huang2020large, gong2021psla, baevski2020wav2vec} in a supervised or unsupervised way, by training models on large amounts of labelled or unlabelled audio. When such models are applied to novel audio, the internal state of the models has been found to contain much of the information necessary for previously unseen tasks. This is demonstrated by training shallow classifiers (probes) on embeddings consisting of the activations of a given model layer, that map these values to a downstream task. Such an approach achieves competitive results using either unsupervised \cite{kim2020one} or supervised \cite{kong2020panns} learning. Most importantly, the embeddings on which the probes are trained are many orders of magnitude smaller than the audio itself and only need to be computed once per audio file. Such embeddings can be stored efficiently, and downstream classifiers can be trained with significantly less resources. The excellent performance, generality, and scalability of this approach are crucial factors for its utility in industry.

This approach to audio understanding has been highlighted in recent benchmarks such as HARES \cite{wang2022towards} and HEAR \cite{HEAR}, where embeddings are evaluated across a number of audio understanding tasks pertaining to a range of content types. Any score aggregation across these benchmarks to determine a "best" embedding is difficult due to the disparate range of metrics employed and furthermore, may obfuscate the strengths and weakness of any given approach. However, evaluating across a common range of tasks can be useful in comparing such strengths and weaknesses. We find that the current tasks evaluated in the HEAR and HARES benchmarks are lacking in evaluation on music content. Wrt. polyphonic music, the HARES benchmark includes only the Magnatagatune dataset, and the HEAR benchmark includes only GTZAN genre and music / speech datasets. While other public music datasets exist, such benchmarks are somewhat limited by the requirement to provide access to the audio of all datasets.

Here, we do not intend to establish a new open benchmark, but investigate the effectiveness of supervised and unsupervised learning for audio embeddings employed specifically for music understanding, across as broad an array of tasks as is available within time and resource constraints. 
For supervised learning we train models on large scale datasets of annotated magnitude log-mel spectrograms both in the music domain and in the general audio domain. For unsupervised learning we train contrastive models using \mbox{SimCLR} loss \cite{chen2020big, chen2020simple} on the same sets of magnitude log-mel spectrograms, excluding annotations.

The contributions of this work are as follows: we provide a broad analysis of supervised and unsupervised learning strategies for pre-training audio models for music understanding; we show that for multilabel / multiclass classification of music, large-scale supervised learning on music data achieves SOTA performance, in many cases outperforming both prior SOTA and unsupervised learning by significant margins; we show that supervised learning on labelled music data does not generalize as well as unsupervised learning to novel tasks not covered in those labels; finally, we show that the domain of pre-training audio datasets has a significant impact on the performance of embeddings, particularly for unsupervised learning.

\section{Pre-training Methodology}\label{sec:pretraining_methodology}

\begin{figure}[t]
\includegraphics[width=8cm]{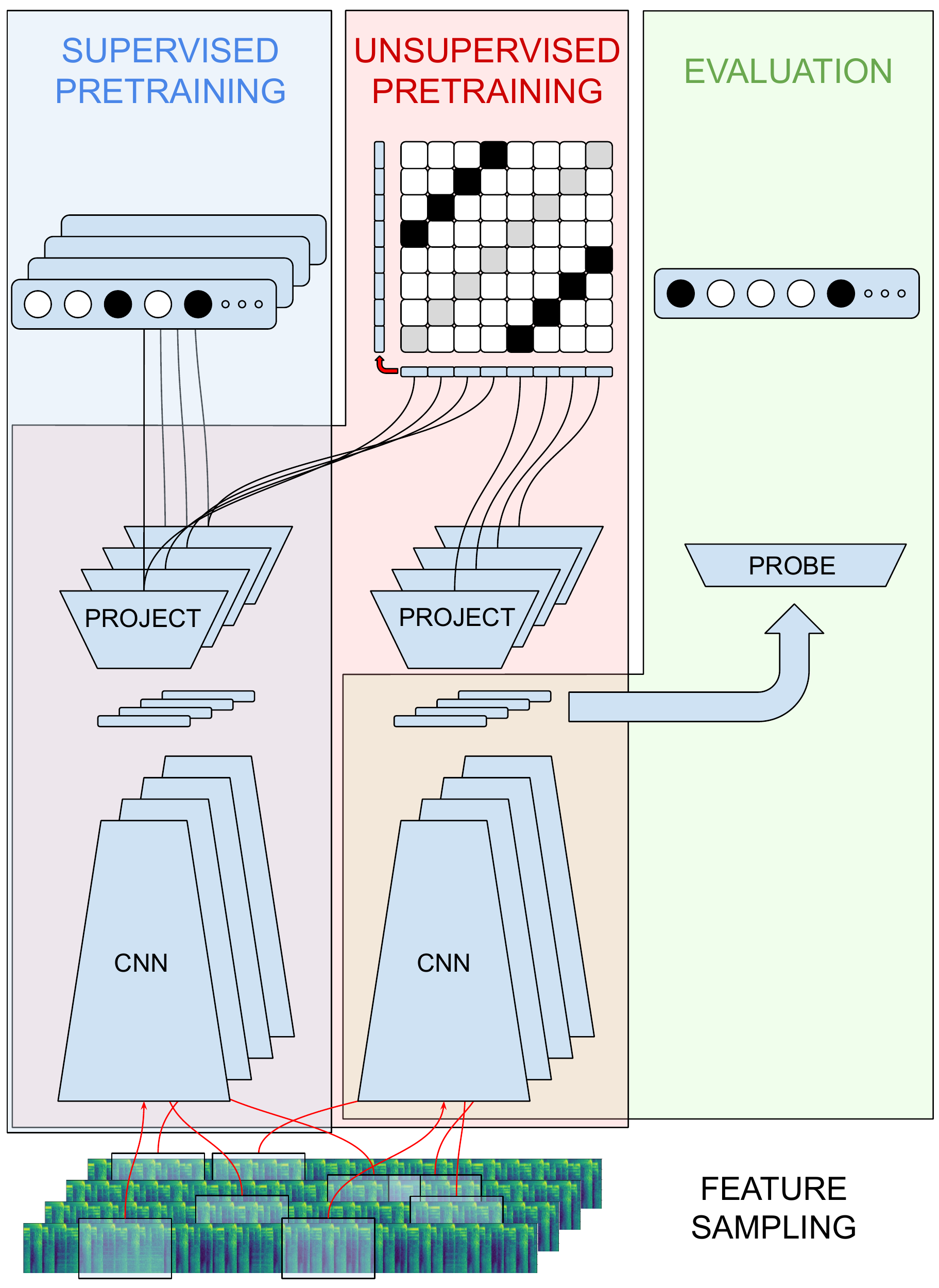}
\caption{System diagram of both pre-training approaches employed in this paper, and evaluation.}
\label{fig:system_diagram}
\end{figure}

To achieve the objectives outlined in Section~\ref{sec:introduction},
we follow a familiar transfer learning paradigm (Figure~\ref{fig:system_diagram}) where models are pre-trained using supervised or unsupervised learning. Thereafter, the frozen activations from a layer of that model, forming embeddings, $\mathbf{z}$, are mapped to a downstream task using a simple network $p(\mathbf{z})$.

\subsection{Supervised and Unsupervised Training}\label{sec:supervised_unsupervised_methods}

In the supervised setting we learn a function $f(\mathbf{X}) \Rightarrow \mathbf{\hat{y}}$ mapping features $\mathbf{X}$ (log-mel spectrograms) to binary labels, $\mathbf{y}$, by applying Adam optimization \cite{kingma2014adam} to the binary cross-entropy loss function,

\begin{align*}
L_s(\mathbf{y}, \mathbf{\hat{y}}) = \frac{-1}{NK} \sum_{i=0}^{N-1} \mathbf{y}_i \log(\mathbf{\hat{y}}_i) + (1-\mathbf{y}_i)\log(1-\mathbf{\hat{y}}_i),
\end{align*}
where batch size $N=512$, and $K$ is the number of labels. 

In the unsupervised setting, we employ the SimCLR objective~\cite{chen2020big, chen2020simple}, which has been shown to provide promising results for both music and audio understanding \cite{wang2022towards, spijkervet2021contrastive}. The SimCLR objective employs correlated (positive) pairs of samples by mapping each feature to an embedding space, $f(\mathbf{X}) \Rightarrow \mathbf{z} \in \mathbb{R}^m$, with embedding dimensionality $m=1728$. A projector, then maps the embedding space to a loss space $h(\mathbf{z}) \Rightarrow \mathbf{v} \in \mathbb{R}^{n}$, with dimensionality $n=1024$. Here, each element is then compared to all other elements in a batch via distance function, $d(\mathbf{v}_i, \mathbf{v}_j)=\mathbf{v}_i \cdot \mathbf{v}_j / \lVert \mathbf{v}_i \rVert \lVert \mathbf{v}_j \rVert$. The loss is then computed as the normalized temperature-scaled cross entropy,

\begin{align*}
   L_u(\mathbf{v}_i,\mathbf{v}_j) = -\log \frac{\exp{\left( d(\mathbf{v}_i, \mathbf{v}_j) / \tau \right)}}{\sum_{k=0}^{2N-1}\mathds{1}_{[k\neq i]} \exp{(d(\mathbf{v}_i, \mathbf{v}_k)/\tau)}},
\end{align*}
which is summed across $2N$ examples in $N=1920$ positive pairs, where $i=j$, for all values of both $i \in [0,N-1]$ and $j \in [0,N-1]$. Here, $\mathds{1}_{[k\neq i]}$ is an indicator function evaluating to 1 for $k=i$, otherwise evaluating to 0. $\tau = 0.1$ denotes a temperature hyper-parameter. 

SimCLR loss can be interpreted as a one hot classification problem---for each example, identifying its positive pair amongst all other (negative) examples in the batch. Because the batch-size determines the number of negative examples in the batch, and hence, the likelihood of non-trivial negatives, a large batch size of $N$ pairs is crucial to this learning strategy, with larger batch sizes resulting in notable performance improvements \cite{wang2021multi}. However, batch sizes are limited to the memory of the available compute resources (e.g., of GPUs / TPUs). Hence, when considering compute costs, it is desirable to reduce batch size.

The primary objective of this work is to provide a comparative analysis of the utility of supervised and unsupervised learning strategies for a range of music understanding tasks under different data sources. As such we do not propose to investigate or innovate on the model architecture defining $f(\mathbf{X})$ itself, and employ the Short-Fast Normalizer-Free Net F0 (SF-NFNet-F0) of \cite{wang2022towards}. This architecture was chosen due to its demonstrated excellent performance in audio understanding, its design intent specifically for audio, and its use of efficient operations such as grouped convolutions. Furthermore, this architecture does not employ batch-normalization, improving training time and resource requirements, particularly when using SimCLR loss, by removing the need to globally synchronize data between GPU devices at each batch-normalization layer. Our intent was to reproduce this model as accurately as possible. However, due to the information available to us at the time of publication, there may be some discrepancies between our implementation and that in \cite{wang2022towards}. To ensure reproducibility, we release a Tensorflow implementation of the model used in this paper with 
SCOOCH configurations\footnote{\url{https://github.com/PandoraMedia/scooch}} 
for each of the models trained\footnote{\url{https://github.com/PandoraMedia/music-audio-representations}}. Our SF-NFNet-F0 implementation contains 62.4M parameters.

In both supervised and unsupervised settings we employ mixup \cite{zhang2017mixup} directly on the sampled log-mel spectrograms in real-time during training for data augmentation. We acknowledge that additively combined audio sources do not result in additively combined magnitude spectrograms due to both constructive and destructive interference. However, such an augmentation is an efficient operation to perform in real-time data sampling pipelines, mitigating data bandwidth bottlenecks that may be caused by more complex operations. We employ mixup by shuffling features within each batch and additively combining the shuffled features (and labels for supervised learning) to the original batch. Mixup gains are sampled from a beta distribution with parameters $\alpha=5.0$ and $\beta=2.0$, for each feature in a batch. 

In all pre-training contexts, we use an Adam optimizer with a learning rate following a warm-up cosine decay schedule, first increasing to 0.0002 over 5k steps, then decreasing to 0.0 over 195k steps. While dense projectors, $h(\mathbf{z})$ are an inherent part of the unsupervised SimCLR learning approach, we find projectors to also be useful in the supervised setting, adding a non-linear transformation between the learned embeddings, $\mathbf{z}$, and the supervised labels, $\mathbf{\hat{y}} = h(f(\mathbf{X}))$. We notice such an approach has also been useful in computer vision \cite{bulent2022improving}. Hence, in all contexts we employ a projector consisting of $3 \times 4096$ node hidden layers with ReLU activation. Supervised models were trained on 8 v100 GPUs taking approximately 30 hours, while unsupervised models were trained on 16 A100 GPUs taking approximately 80 hours.

\subsection{Datasets}\label{sec:trainingdata}

Pre-training datasets can have a significant impact on supervised and unsupervised model performance. In the supervised context, there has been evidence that models are less generalizable to unseen tasks \cite{wang2022towards}, perhaps due to the information present in the supervised labels. In the unsupervised context, less investigation has been conducted into the effect of the content of pre-training datasets. In this context, the problem of sampling positive and non-trivial negative examples within a dataset is closely tied to the diversity of content in that dataset. Furthermore, when employing mixup as a data augmentation strategy, the content of the pre-training dataset also defines the additive noise that is applied to features during training.

We compare pre-training on two large datasets, namely \textbf{Musicset} and a version of \textbf{Audioset} \cite{45857}. The Audioset training set contains $\approx$ 1.7M distinct labelled content pieces, and Musicset $\approx$ 1.8M. Audioset consists of 10~second snippets of various audio sources: music, speech and environmental audio (4,791 hours, $\approx$ 602~GB of audio feature data). Musicset, in contrast, focuses solely on music; it consists only of labelled complete songs, each up to several minutes in length (117,497 hours, $\approx$ 14.7~TB of audio feature data). Musicset is, to our knowledge, the largest dataset of expert-annotated audio ever trained on. While it is not publicly available, we believe it valuable to report the results of models trained on this dataset to communicate the effectiveness of supervised learning at this scale.

For supervised learning, in addition to the features themselves, the labels differ. Each dataset's vocabulary is distinct but similar in size (527 labels for Audioset and 500 labels for Musicset), however, Figure~2 shows the label density and distribution of both datasets differs.

\subsection{Feature Sampling}\label{Feature Sampling}

The features, $\mathbf{X}$, are log-magnitude log-mel spectrograms produced from waveforms sampled at 16~kHz. We use 96 HTK-log-mel spaced, power normalized, frequency bins with center frequences from 0~Hz to 8~kHz, analyzed with a window size of 25~ms, a Fourier transform size of 2048 bins and a hop size of 10~ms. We sample 3~s snippets of each content piece in real-time (during training) across the pre-training dataset, forming features, $\mathbf{X} \in \mathbb{R}^{96 \times 300}$. Sampling features in real-time from full-length tracks has the benefit of a high ratio of distinct features per static dataset size by reducing redundant (overlapping) data storage. For example, a 183~second song (6.8MB) results in 18k distinct features (each 112kB) with the above parameters.

To build a batch from a dataset, we first select a uniform random sample of tracks (with replacement); then, from each selected track, we choose a random 3-second spectrogram snippet, without padding. In the unsupervised context, we need to select positive and negative pairs for each selected snippet. Positive pairs are sampled from 
the same 10~second "track" for Audioset; for Musicset, positive examples are also sampled from the same track, although we require positive pairs to be centered on frames less than 5~seconds apart in the track's timeline. As negative examples, the SimCLR objective implicitly uses all other samples in a batch. We also tried forcing examples from the same track, but further away from the anchor than the positive, into every batch as hard negatives. Such an approach has shown promising results in music segmentation \cite{mccallum2019unsupervised}, however, we found that for music labelling at the averaged track-level this did not change our results significantly. As such, we omit this approach from our evaluation.

\begin{figure}[t]
\label{fig:label_distribution}
\includegraphics[width=8cm]{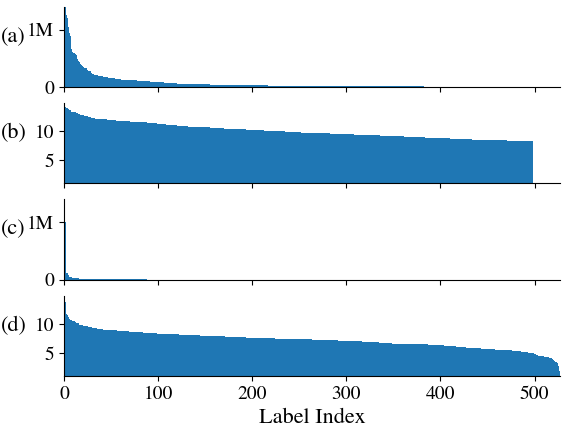}
\caption{Label density distributions for Musicset and Audioset. (a), (b) plots the sorted Musicset label counts on a linear (in millions) and logarithmic scale, respectively. (c), (d) plots the same data for Audioset, respectively.}
\end{figure}

\section{Evaluation}\label{sec:evaluationdata}

To evaluate embedding performance, we take the global average pooled activations of the final convolutional layer in the SF-NFNet-F0 architecture for 3~second feature snippets sampled along the length of each audio timeline at a frequency of 0.5~Hz, then average these sampled embeddings along the length of the timeline (with the exception of NSynth datasets as described later in this section). We found similar results training probes on both the timeline-averaged embeddings, and individual embeddings sampled directly from track timelines, with slight improvements when using timeline-averaged embeddings in some cases.

Each of the embeddings derived from pre-trained models have different levels of predictive power and will underfit / overfit with different parameters for each downstream-dataset / pre-trained model combination. To demonstrate the potential of embeddings, and to investigate whether a given embedding can achieve SOTA performance on a given dataset, it is important to optimize the probe parameters for each dataset. However, in cases where such a transfer learning methodology is SOTA, it is important to constrain the probe to the limitations imposed in previous works so that it is the quality of embeddings that are evaluated, and not that of the probes. With these considerations, we adopt a hybrid strategy based on the prior SOTA for each dataset. We optimize the parameters of probes in all cases except where the prior SOTA adopts a similar transfer learning methodology, in which case, we constrain the probes to the parameter ranges investigated in those prior works. Regardless of prior works, we restrict all probes to simple multi-layer perceptron (MLP) classifiers which can be trained on a single 32~core CPU in less than 1~hour to maintain the aforementioned benefits of efficiency and scalability that such a transfer learning approach provides.

In all cases, we train probes using Adam optimization with a cosine learning rate schedule with 1,000 steps of warmup followed by a decay to zero over the remainder of the steps. We optimize the learning rate, number of training steps, and l2-regularization of each probe to achieve best performance / prevent overfitting. 

In total we evaluate the performance over annotations of 7 distinct collections of audio, comprising 15 distinct datasets in total. An overview of datasets is shown in Table~\ref{table:datasets}, and probe configurations in Table~\ref{table:probes}. We also publish more detailed SCOOCH configurations for probes, tag-level results, and model weights for the Musicset-ULarge model as a baseline for future research.\footnote{\url{https://github.com/PandoraMedia/music-audio-representations}} The remainder of this section covers specific notes on each dataset.

\begin{table}[ht!]
  \centering
  \footnotesize
  \begin{tabular}{@{}cccccc@{}}
    \textbf{Dataset} & \textbf{Task} & \textbf{\#Items} & \textbf{\#Labels} & \textbf{Avg. secs} & \textbf{Hours}\\
    \toprule
    MSDS & tagging & 242k & 50 & 46 & 3.08k \\
    MSD50 & tagging & 36k & 50 & 46 & 464 \\
    MSD100 & tagging & 115k & 100 & 47 & 1.50k \\
    MSD500 & tagging & 156k & 500 & 47 & 2.05k \\
    AMM & mood & 67k & 188 & 45 & 840 \\
    MuMu & genre & 147k & 250 & 47 & 1.92k \\
    MTT & tagging & 26k & 50 & 29 & 171 \\
    NSynth\textsubscript{P} & pitch & 306k & 112 & 4 & 340 \\
    NSynth\textsubscript{I} & instrument & 306k & 11 & 4 & 340 \\
    GTZAN & genre & 930 & 10 & 30 & 7.8 \\
    Emo & emotion & 744 & N/A & 45 & 9.3 \\
    GS\textsubscript{Key} & key & 2.1k & 24 & 120 & 116 \\
    Jam-50 & tagging & 54k & 50 & 244 & 3.69k \\
    Jam-All & tagging & 56k & 183 & 244 & 3.76k \\
    Jam-MT & mood/theme & 18k & 56 & 219 & 1.10k \\
    \bottomrule
  \end{tabular}
  \caption{Overview of evaluation datasets. The sizes for MSD50 and MSD500 disagree with \cite{won2021multimodal}. The published MSD50 splits include 35,745 IDs, and we exclude any tracks in MSD500 without any tags in the vocabulary.}
  \label{table:datasets}
\end{table}

\begin{table}[ht!]
  \centering
  \footnotesize
  \begin{tabular}{@{}ccccc@{}}
    \textbf{Dataset} & \textbf{Layers} & \textbf{Dropout} & \textbf{Batch ($N$)} & \textbf{Constraint} \\
    \toprule
    MSDS & 2$\times$1024 & 0.5 & 256 & None \\
    MSD50 & 1$\times$512 & 0.5 & 256 & None \\
    MSD100 & 2$\times$2048 & 0.5 & 256 & None \\
    MSD500 & 3$\times$4096 & 0.5 & 256 & None \\
    AMM & 1$\times$1024 & 0.5 & 256 & None \\
    MuMu & linear & 0.3 & 256 & None \\
    MTT & 1$\times$512 & 0.5 & 256 & \cite{castellon2021codified} \\
    NSynth\textsubscript{P} & linear & 0.0 & 64 & \cite{wang2022towards} \\
    NSynth\textsubscript{I} & linear & 0.0 & 64 & \cite{wang2022towards} \\
    GTZAN & linear & 0.0 & 2560 & None \\
    Emo & 1$\times$1024 & 0.5 & 512 & None \\
    GS\textsubscript{Key} & 1$\times$512 & 0.8 & 512 & None \\
    Jam-50 & 1$\times$512 & 0.75 & 256 & None \\
    Jam-All & 2$\times$1024 & 0.5 & 256 & None \\
    Jam-MT & 1$\times$512 & 0.75 & 256 & None \\
    \bottomrule
  \end{tabular}
  \caption{Probe parameters}
  \label{table:probes}
\end{table}

\textbf{Magnatagatune:} Magnatagatune (MTT) \cite{law2009evaluation} annotations include genre, instrumentation / vocal, mood, perceptual tempo, origin and sonority features.
We adopt the common approach using the published splits and top 50 tags\footnote{\label{jpl_splits}\url{https://github.com/jongpillee/music_dataset_split}}. 
Some research removes tracks without tags in the most common 50 tags.
To make results comparable to the most recent SOTA on this dataset \cite{castellon2021codified}, we keep items without tags in both training and evalution sets. 

\textbf{GTZAN:} This dataset addresses the problem of genre classification in a single-label context  \cite{tzanetakis2002musical}. 
We employ the fault-filtered version of this dataset\footnotemark[\getcountref{jpl_splits}].

\textbf{NSynth:} Here we evaluate detecting the pitch (NSynth\textsubscript{P}) and instrument (NSynth\textsubscript{I}) of individual musical notes. Following \cite{wang2022towards}, in the case of NSynth\textsubscript{P} we analyze the average of embeddings from four non-overlapping consecutive snippets of 1~second feature windows, and for NSynth\textsubscript{I} we analyze a single embedding for each note produced using a 4~second feature window.

\textbf{Million Song Dataset:} The Million Song Dataset (MSD) 
contains labels pertaining to era, instrumentation, sonority, genre, mood, origin and activity. Because of the size of this dataset and variation in label quality throughout we adopt several different splits and vocabularies for the dataset. For comparability with the previous SOTA, we adopt the vocabulary of the 50 most common labels, and the splits employed in \cite{won2021semi}\footnotemark[\getcountref{jpl_splits}],
namely MSDS. 
We also take advantage of the several cleaned and artist-separated datasets available for this collection \cite{won2021multimodal}---MSD50, MSD100 and MSD500, for which the
splits and vocabularies 
are available publicly\footnote{\url{https://github.com/minzwon/tag-based-music-retrieval}}. 

\textbf{All Music Moods:} The All Music Moods dataset (AMM) \cite{korzeniowski2020mood}, focuses on mood prediction for the MSD audio data. Because the mood labels are heavily correlated with the artists in the dataset, we find it important to adopt the artist-seperated split\footnote{\url{https://github.com/fdlm/listening-moods}}. With no previously published result on this split, we additionally provide results for embeddings produced by the MusiCNN model \cite{pons2019musicnn}. 

\textbf{Jamendo:} This dataset contains genre, instrument and mood / theme tags for audio from Jamendo \cite{bogdanov2019mtg}. We evaluate on all tags (Jam-All) and the 50 most common (Jam-50). Because very few of the mood / theme tags are in Jam-50, we also evaluate the mood / theme category (Jam-MT). We use the official splits\footnote{\url{https://github.com/MTG/mtg-jamendo-dataset}} and full-length audio. 

\textbf{GiantSteps Key:} This dataset (GS\textsubscript{Key}) concerns major/minor key classification in electronic music---a 24-way classification problem. It combines two 
datasets: the first, 604 2-minute samples of electronic music tracks 
collected from \url{beatport.com}~\cite{knees_two_2015}; the second\footnote{\url{https://github.com/GiantSteps/giantsteps-mtg-key-dataset}}, consists of 1486 tracks from the same source. Consistent with previous work~\cite{dhariwal2020jukebox,korzeniowski_end--end_2017}, we use the former 604 samples for testing, and the latter for training / validation, selecting high-confidence annotations
partitioned into train and validation sets according to~\cite{dhariwal2020jukebox}. For evaluation we compute a weighted accuracy score common in key classification\footnote{\url{https://www.music-ir.org/mirex/wiki/2021:Audio_Key_Detection}}.

\textbf{EmoMusic:} This dataset (Emo) concerns emotion recognition~\cite{soleymani_1000_2013}, and provides
continuous annotations for valence and arousal. Following~\cite{dhariwal2020jukebox}, we consider the static version of this dataset, where target values are averaged for each clip, and use the artist-based train / test partitioning provided in their work. This poses a regression problem---we use the coefficient of determination as the evaluation metric for both valence (Emo\textsubscript{V}) and arousal (Emo\textsubscript{A}). 

\textbf{MuMu:} The Multimodal Music dataset (MuMu)~\cite{oramas2018multimodal}, focuses on multilabel genre predictions for the MSD audio data, with genre annotations from the Amazon Reviews dataset. We evaluate 
using the official splits\footnote{\url{https://zenodo.org/record/1236906\#.YoPIAhNBx0s}}.

\section{Results}\label{sec:results}

\begin{table*}[ht!]
  \centering
  \footnotesize
  \begin{tabular}{@{}ccccccccccccc@{}}
    \multirow{2}{*}{\textbf{Model}}& \multicolumn{2}{c}{MTT} & GTZAN & NSynth\textsubscript{P} & NSnyth\textsubscript{I} & Emo\textsubscript{V} & Emo\textsubscript{A} & GS\textsubscript{Key} & \multicolumn{2}{c}{Jam-50} & \multicolumn{2}{c}{Jam-All} \\
     & mAP & ROC & Acc & Acc & Acc & r\textsuperscript{2} & r\textsuperscript{2} & W. Acc & mAP & ROC & mAP & ROC \\
    \toprule[1.1pt]
    Musicset-Sup & \textbf{0.413} & \textbf{0.917} & \textbf{0.835} & 0.793 & 0.731 & 0.566 & \textbf{0.726} & 0.286 & \textbf{0.321} & \textbf{0.843} & \textbf{0.162} & \textbf{0.839} \\
    Audioset-Sup & 0.386 & 0.904 & 0.748 & 0.819 & 0.676 & 0.341 & 0.545 & 0.210 & 0.284 & 0.822 & 0.135 & 0.813 \\
    \midrule
    Musicset-ULarge & \textbf{0.404} & \textbf{0.914} & 0.735 & \textbf{0.892} & 0.740 & 0.577 & 0.700 & 0.667 & \textbf{0.317} & \textbf{0.839} & \textbf{0.159} & \textbf{0.833} \\
    Audioset-ULarge & 0.391 & 0.906 & 0.672 & 0.805 & 0.721 & 0.438 & 0.624 & 0.287 & 0.285 & 0.826 & 0.131 & 0.816 \\
    \midrule
    Musicset-USmall & 0.389 & 0.905 & 0.686 & 0.824 & 0.714 & 0.389 & 0.668 & 0.508 & 0.292 & 0.828 & 0.138 & 0.817 \\
    Audioset-USmall & 0.375 & 0.897 & 0.648 & 0.777 & 0.698 & 0.386 & 0.609 & 0.197 & 0.268 & 0.817 & 0.127 & 0.809 \\
    \midrule[1.1pt]
    Jukebox \cite{dhariwal2020jukebox} & \textbf{0.414} & \textbf{0.915} & 0.797 & - & - & \textbf{0.617} & \textbf{0.721} & 0.667 & - & - & - & - \\
    Prev. SF-NFNet-F0 \cite{wang2022towards} & 0.395 & - & - & 0.880 & \textbf{0.782} & - & - & - & - & - & - & - \\
    \midrule
    \multirow{2}{*}{SOTA Excl. \cite{dhariwal2020jukebox, wang2022towards}} & 0.384 & \textbf{0.92} & 0.821 & - & 0.741 & 0.556 & 0.704 & \textbf{0.796}* & 0.298 & 0.832 & - & - \\
                            & \cite{pons2019musicnn} & \cite{huang2020large} & \cite{lee2018samplecnn} & - & \cite{niizumi2021byol} & \cite{koh2021comparison} & \cite{weninger2014line} & \cite{rekordbox} & \cite{won2020evaluation} & \cite{won2020evaluation} & - & - \\
    \bottomrule[1.1pt]
    \end{tabular}  
    \caption{Results for models on all datasets, excluding MSD annotations and Jam-MT. Models within 0.01 of SOTA are bold. *The SOTA for GS\textsubscript{Key} has no publicly available implementation details and is specialized for this dataset, e.g., \cite{bernardes2017automatic}.}
  \label{table:nonmsd}
\end{table*}

\begin{table*}[ht]
  \centering
   \footnotesize
  \begin{tabular}{@{}ccccccccccccccc@{}}
    \multirow{2}{*}{\textbf{Model}} & \multicolumn{2}{c}{MSDS} & \multicolumn{2}{c}{MSD50} & \multicolumn{2}{c}{MSD100} & \multicolumn{2}{c}{MSD500} & \multicolumn{2}{c}{MuMu} & \multicolumn{2}{c}{AMM} & \multicolumn{2}{c}{Jam-MT} \\
     & mAP & ROC & mAP & ROC & mAP & ROC & mAP & ROC &  mAP & ROC &  mAP & ROC &  mAP & ROC \\
    \toprule[1.1pt]
    Musicset-Sup & \textbf{0.363} & \textbf{0.903} & \textbf{0.459} & \textbf{0.913} & \textbf{0.346} & \textbf{0.906} & \textbf{0.169} & \textbf{0.898} & \textbf{0.257} & \textbf{0.908} & \textbf{0.180} & \textbf{0.791} & \textbf{0.161} & \textbf{0.786} \\
    Audioset-Sup & 0.308 & 0.880 & 0.375 & 0.883 & 0.278 & 0.877 & 0.128 & 0.874 & 0.191 & 0.867 & 0.156 & 0.760 & 0.137 & 0.749 \\
    \midrule
    Musicset-ULarge & 0.351 & \textbf{0.900} & 0.438 & \textbf{0.908} & 0.321 & \textbf{0.897} & 0.152 & \textbf{0.891} & 0.235 & 0.893 & \textbf{0.174} & \textbf{0.784} & \textbf{0.158} & \textbf{0.781} \\
    Audioset-ULarge & 0.311 & 0.885 & 0.377 & 0.886 & 0.276 & 0.878 & 0.121 & 0.873 & 0.162 & 0.855 & 0.156 & 0.763 & 0.142 & 0.765 \\
    \midrule
    Musicset-USmall & 0.319 & 0.888 & 0.384 & 0.892 & 0.283 & 0.881 & 0.129 & 0.878 & 0.190 & 0.871 & 0.155 & 0.762 & 0.138 & 0.757 \\
    Audioset-USmall & 0.286 & 0.876 & 0.353 & 0.878 & 0.251 & 0.870 & 0.110 & 0.868 & 0.152 & 0.850 & 0.151 & 0.753 & 0.136 & 0.753 \\
    \midrule[1.1pt]
    \multirow{2}{*}{SOTA}  & 0.348 & \textbf{0.897} & 0.386 & \textbf{0.921} & 0.185  & - & - & - & - & 0.888* & 0.163 & 0.773 & \textbf{0.161}$^\dagger$ & \textbf{0.781}$^\dagger$ \\ 
                           & \cite{won2021semi} & \cite{won2021semi} &\cite{lu2021spectnt} & \cite{lu2021spectnt} & \cite{won2021multimodal} & - & - & - & - & \cite{oramas2018multimodal} & \cite{pons2019musicnn} & \cite{pons2019musicnn}      & \cite{knox2020mediaeval} & \cite{knox2020mediaeval} \\ 
    \bottomrule[1.1pt]
  \end{tabular}
  \caption{Results for all models on MSD annotations and Jam-MTT. Models within 0.01 of SOTA are bold. *The previous SOTA for MuMu evaluated artist-level predictions rather than track-level. $^\dagger$The SOTA for Jam-MT is trained on an expanded within-taxonomy set ($\approx$ 16k tracks), here we restrict probe training to Jamendo tracks ($\approx$ 10k).}
  \label{table:msd}
\end{table*}

In order to derive conclusions on the potential performance of supervised models trained using binary labels on music data, we train a supervised model on the complete Musicset dataset, namely \emph{Musicset-Sup}. To compare these results to supervised learning on a large-scale public dataset covering music, environmental, and speech audio, we provide results for a model trained on $\approx$1.7M items from Audioset, namely \emph{Audioset-Sup}. To compare supervised and unsupervised learning at a similar scale, we provide results for a model trained using the unsupervised methodology discussed in Section~\ref{sec:pretraining_methodology} on the same Musicset dataset audio data and the Audioset audio data, namely \emph{Musicset-ULarge} and \emph{Audioset-ULarge} using a batch size of 1920 pairs.

In addition, we train two further unsupervised models using a batch-size of 256. One trained on Audioset (\emph{Audioset-USmall}), and one on $\approx$1.8M randomly sampled 10~second snippets from Musicset, one per content piece (\emph{Musicset-USmall}). The reasoning for this is two-fold---firstly by sampling short snippets from Musicset we mitigate the confounding variable of dataset size when comparing the two, secondly this allows us to investigate the effect of batch size in the unsupervised learning paradigm.

For convenience, we also include results for the recent evaluations in \cite{wang2022towards, castellon2021codified}, and those for any other previous SOTA, excluding these two evaluations. The results for each of these models is shown in Table~\ref{table:nonmsd} for all datasets excluding annotations of MSD, which are shown in Table~\ref{table:msd}. There, we note that for all multilabel music tagging tasks, the Musicset-Sup model achieves SOTA performance by significant margins. This is encouraging given that the Musicset training dataset was created naively, and the supervised information therein opens the door to improvements such as label-balanced sampling which has been shown to realize further performance gains \cite{kong2020panns}. We leave this investigation for future work.

It is also encouraging to see that unsupervised learning on music audio (Musicset-ULarge) achieves SOTA performance compared to previous unsupervised models, in addition to some supervised models (specifically, for models trained on Jam-50 and all MSD datasets). 

Comparing to recent transfer learning approaches~\cite{wang2022towards,castellon2021codified}, we see that in all cases either Musicset-Sup or Musicset-ULarge outperform or are on par with these, with the exception of NSynth\textsubscript{I}. This is a promising result espousing the value of music only audio datasets considering that previously reported results for unsupervised learning on Audioset \cite{wang2022towards} used more than double the batch size of those in this paper. Wrt.\ the Jukebox model, models trained for this paper use approximately 1\% of the parameters and take less than 1\% of the GPU flop hours to train.

We see that the supervised models evaluated demonstrate shortcomings in performance on pitch (NSynth\textsubscript{P}) and key (GS\textsubscript{Key}) tasks. This corroborates the findings in \cite{castellon2021codified} that models trained on tags do not perform well in this task, and the findings in \cite{wang2022towards} that supervised pre-trained models do not generalize as well to novel tasks. We note that there are no pitch or key labels in the Musicset dataset, and in fact, many of the labels employed (e.g., genre, mood, etc.) require the model to be somewhat agnostic to such information. It is yet to be seen whether including such information in the labels of a pre-training dataset would improve the generalizability of such models, though this is outside of the scope of the current work. Interestingly, we see that unsupervised models trained specifically on music data show significant improvements in key estimation.

When comparing the domain of the pre-training dataset features in the models of Musicset-USmall and Audioset-USmall we see that in all cases, in-domain data (music) results in a better performing model. We conjecture that this may be due to (a) increasing the chances of non-trivial negative examples within each batch relative to each positive pair, due to the homogeneity of the dataset, and (b) music is more spectrally dense and correlated than other common noise sources employed in mixup augmentation, such as speech or various environmental noises.

\section{Conclusions}\label{sec:conclusion}

In this work we investigated both unsupervised and supervised learning strategies, to produce compact audio representations that may be deployed across industry-scale audio catalogs for a range of downstream use cases. We observed that supervised training on large scale expert-annotated music data achieves SOTA results in music tagging. We see that unsupervised learning on datasets of the same scale also achieves excellent performance, across a wider range of tasks than supervised learning, particularly on those that involve information not represented in the supervised dataset. Finally, we see that for unsupervised learning, restricting the domain of the pre-training dataset to music results in improvements in model performance.

\section{Acknowledgements}\label{sec:acknowledgements}

The authors would like to thank the Pandora music analyst team for their years of dedicated and careful work that enables us to investigate supervised machine learning systems at modern levels of quality and scale.

\bibliography{ismir2022}

\end{document}